\documentclass[%
reprint,
superscriptaddress,
showpacs,preprintnumbers,
amsmath,amssymb,
prb,
]{revtex4-1}

\usepackage{graphicx}% Include figure files
\usepackage{dcolumn}% Align table columns on decimal point
\usepackage{bm}% bold math
\usepackage{color}

\begin{document}

\preprint{APS/123-QED}

\title{
Nature of the $a_1$ meson in lattice quantum chromodynamics studied with chiral fermions
}

\author{Yuko~Murakami}%
\affiliation{Research and Development Laboratory, Seikow Chemical Engineering $\&$ Machinery, LTD, Akashi 674-0093, Japan}
\author{Shin~Muroya}%
\affiliation{Matsumoto University, Matsumoto 390-1295, Japan}
\author{Atsushi~Nakamura}%
\affiliation{School of Biomedicine, Far Eastern Federal University, 690950 Vladivostok, Russia}
\affiliation{Theoretical Research Division, Nishina Center, RIKEN, Wako 351-0198, Japan}
\affiliation{Research Center for Nuclear Physics (RCNP), Osaka University, Ibaraki, Osaka 567-0047, Japan}
\author{Chiho~Nonaka}%
\affiliation{Department of Physics, Nagoya University, Nagoya 464-8602, Japan}
\affiliation{Kobayashi Maskawa Institute, Nagoya University, Nagoya 464-8602, Japan}
\author{Motoo~Sekiguchi}%
\affiliation{School of Science and Engineering, Kokushikan University, Tokyo 154-8515, Japan}
\author{Hiroaki~Wada}%
\affiliation{School of Science and Engineering, Kokushikan University, Tokyo 154-8515, Japan}
\author{Masayuki~Wakayama}%
\affiliation{Research Center for Nuclear Physics (RCNP), Osaka University, Ibaraki, Osaka 567-0047, Japan}

\collaboration{SCALAR Collaboration}%\noaffiliation

\date{\today}% It is always \today, today,

\begin{abstract}
We study the $a_1$ meson using a quenched lattice quantum chromodynamics simulation 
with the truncated overlap fermions formalism 
based on the domain wall fermions. 
The obtained lightest mass of the $a_1$ meson, 1272(45)\ MeV, is consistent with the experimental value for $a_1$(1260). 
Thus, $a_1$(1260) can be identified to have a simple two-body constituent-quark structure. 
Our quenched simulation result of $a_1$(1420) can not explain the experimental mass value, 
which suggests $a_1$(1420) is not a simple $q\bar{q}$ two quark state.


%\begin{description}
%\item[PACS numbers] \pacs{12.38.Gc,  14.40.Be,   14.40.Rt,     12.40.Yx}
%,  14.40.Be,   14.40.Rt,     12.40.Yx}
%\end{description}
\end{abstract}

%\pacs{12.38.Gc,  14.40.Be,   14.40.Rt,     12.40.Yx}% PACS, the Physics and Astronomy
\maketitle

%\tableofcontents

%\section{\label{sec:level1}First-level heading:\protect\\ The line
%break was forced \lowercase{via} \textbackslash\textbackslash}

%\section{Introduction}
\noindent\underline{\bf Introduction}
\vspace{0.5mm}

In hadron spectroscopy, the fundamental ingredients are 
light-meson sector, whose understanding both from 
the theoretical and experimental aspects are indispensable. 
And yet, the classification of light axial-vector mesons ($a_1$) is a long-standing issue in the meson spectroscopy. 
Recently, the resonance of $a_1$(1260) was observed clearly in the COMPASS experiment at CERN~\cite{COMPASS:2010}. 
Moreover, a new $a_1$ meson was discovered in the $f_0$$\pi$ channel 
with a mass of 1420\ MeV and a narrow width by the COMPASS collaboration~\cite{COMPASS:2015}. 
Currently the particle data group lists three $a_1$ mesons~\cite{PDG2018}: 
$a_1$(1260), $a_1$(1420), and $a_1$(1640); 
however, 
this is a richer spectrum than that in the usual $q\bar{q}$ mesons in a constituent quark model. 
In the conventional constituent quark model~\cite{PDG2018}, 
the $a_1$(1260) meson is assigned to a $I=1$, $^{3}P_{1}$ state. 
If $a_1$(1260) is the ground state for $a_1$ meson, the mass of the next radial excitation becomes 
at least 1.7\ GeV~\cite{Ebert:2009ub}. 
This suggests that the radial excitation of $a_1$(1260) can not be $a_1$(1420),  but $a_1$(1640). 
Consequently, the structure of $a_1$(1420) cannot be understood as a simple two-quark state. 
It is a possible candidate for the exotic multi-quark state~\cite{Chen, Gutsche2017, Gutsche2018, Sundu} 
or the dynamical effect due to a singularity in the triangle diagram~\cite{Mikhasenko, Aceti, Basdevant, Wang:2016}. 

There have been several interpretations 
for the structure of the $a_1$(1260) meson: 
i) in the Nambu-Jona-Lasinio model~\cite{Nambu}, 
the $a_1$(1260) meson is the chiral partner of the $\rho$ meson 
as $q\bar{q}$ state, 
ii) it could also be interpreted 
as the gauge boson of the hidden local symmetry~\cite{Bando:1985, Yamawaki, Bando:1988,Kaiser},
iii) in the coupled-channel approaches based on chiral effective theory~\cite{Roca}, 
it is described as the dynamically generated resonance in $\pi$$\rho$ scattering,
and
iv) Nagahiro {\it et al.} discussed the mixing properties of $a_1$(1260) of the quark composite state 
and the hadronic composite state~\cite{Nagahiro}. 

In this report, 
we present the structure of the lightest $a_1$ meson determined with lattice 
quantum chromodynamics (QCD),  a first-principles approach. 
Our objective is to clarify 
the relation between the nature of the $a_1$ meson 
and the chiral symmetry associated with the chiral partner of the $\rho$ meson
and dynamical chiral symmetry breaking,  
as is the case for $\pi$ and the chiral partner of the $\sigma$ meson~\cite{Nambu}.
We, therefore, employ the truncated overlap fermion formalism 
by Bori\c{c}i~\cite{Borici:1999zw} 
based on domain wall fermions formalism~\cite{Kaplan:1992bt,Furman:1994ky}, 
which holds good chiral symmetry. 
The truncated overlap fermion formalism is classified into lattice chiral fermions
~\cite{Kaplan:1992bt,Furman:1994ky,Narayanan:1993ss,Brower:2004xi}. 

In the previous work~\cite{Scalar:2004}, 
we investigated the $\sigma$ meson 
based on the full QCD with dynamical Wilson quarks that has an explicit chiral symmetry
breaking term, 
using $q\bar{q}$ interpolating operators. 
Our work indicates the existence of the light $\sigma$ state,  
whose mass is in $m_{\pi} < m_{\sigma} \leq m_{\rho}$. 
The disconnected diagram plays an essential role in the $\sigma$ meson mass becoming small. 
Unlike the $\sigma$ meson, the $a_1$ meson propagator does not have a disconnected diagram. 
Thus, the quenched lattice simulation 
is able to show 
whether the $a_1$ meson can fit the simple constituent quark model. 

Lattice simulations of the $a_1$ meson have been previously conducted. 
Wingate {\it et al.} were the first to measure the mass of the $a_1$ meson using lattice QCD 
with two flavors of dynamical staggered quarks~\cite{Wingate:1995hy}. 
Their result was in agreement with the experiment value for $a_1$(1260). 
Recently, Gattringer {\it et al.} determined the mass of the $a_1$ meson 
utilizing the chirally improved Dirac operator in the quenched approximation 
with the L${\rm \ddot{u}}$scher-Weisz gauge action~\cite{Gattringer:2008}. 
These simulations demonstrated a clear improvement in the description of the ground state 
employing interpolators with derivative quark sources. 
Their obtained mass of the ground state $a_1$ meson is close to $a_1$(1420), 
instead of $a_1$(1260), i.e., these simulations are inconsistent. 

Here, we perform quenched simulations for the $a_1$ meson 
using the truncated overlap fermion formalism with $q\bar{q}$ interpolating operators. 
We will show that the lightest $a_1$ meson is the $q\bar{q}$ state composed of $u$ and $d$ quarks. \\

\begin{figure}%[h]
\begin{center}
\includegraphics[scale=0.48]{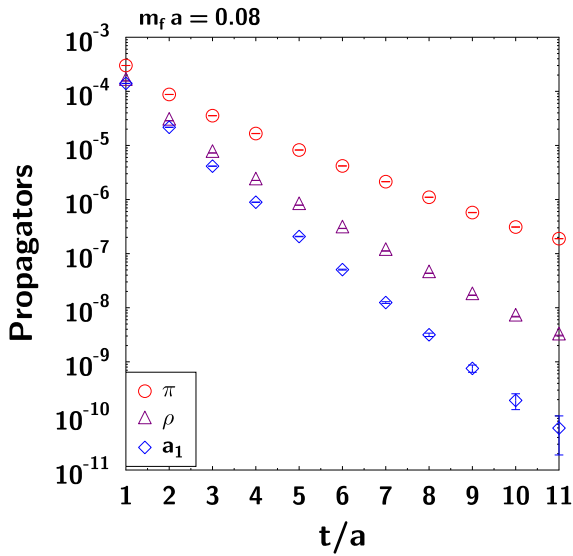} \ \ \ 
\caption{ (color online). 
Time dependence of the propagators at $m_f a = 0.08$. 
Open circles, triangles, and diamonds represent the propagators of $\pi$ meson, $\rho$ meson, and $a_1$ meson, respectively. 
}
\label{propa}
\end{center}
\end{figure}

%\section{Lattice simulation}
\noindent\underline{\bf Lattice simulation}
\vspace{0.5mm}

We perform quenched lattice QCD calculations using truncated overlap fermions~\cite{Borici:1999zw} 
with the plaquette gauge action. 
We use point sources and sinks when calculating hadron propagators, 
which 
leads to larger masses on a relatively small lattice 
because of a mixture of higher mass states. 
The masses obtained in our simulation should thus be considered as the upper limits. 
The $a_1$ meson propagator is more noisy than those of $\pi$ and $\rho$ mesons, 
and therefore more statistics are required. 
Since truncated overlap fermions are a variant of domain wall fermions, 
we use the same simulation parameters as those used by Blum {\it et al.}~\cite{Blum:2000kn}, 
except for the temporal lattice size ($N_t = 24$ is here used, instead of $N_t = 32$.): 
$\beta=5.7$, the length of the fifth dimension $N_{5}=32$ for which $m_{\pi}^2$ is stable, 
the five-dimensional mass $m_{5} = 1.65$, 
and 
the three-dimensional spatial lattice size $N_s^3=8^3$. 

We adopt the following interpolating operator for creating the $a_1$ meson with $I=1$ and $J^{PC}=1^{++}$, 
\begin{eqnarray}
O_{a_1} = \bar{q} \gamma_{\mu} \gamma_5 q \ ,
\end{eqnarray}
where $q$ denotes the $u$ or $d$ quark operator. 
We generate gauge configurations based on the plaquette gauge action by using the pseudo heat-bath method. 
After 20000 thermalization iterations, we start to save gauge configurations every 1000 sweeps. 
We calculate meson propagators on the stored gauge configurations 
for each of the quark mass values, $m_f a = 0.08$, 0.06, and 0.04, where $a$ is the lattice spacing. 
We use 3000 (7964) configurations for the calculation of the meson propagators 
with $m_f a =0.08$ and 0.06 ($m_f a = 0.04$). 

\begin{figure}%[h]
\begin{center}
\includegraphics[scale=0.48]{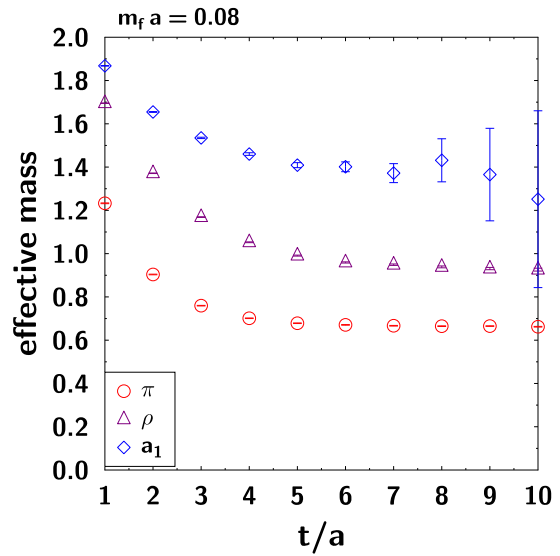} \ \ \ 
\caption{ (color online). 
Time dependence of the effective masses at $m_f a =0.08$. 
Open circles, triangles, and diamonds represent the propagators of $\pi$ meson, $\rho$ meson, and $a_1$ meson, respectively. 
}
\label{eff_mass}
\end{center}
\end{figure}

\begin{table}[b]
\caption{
Masses of $\pi$, $\rho$, and $a_1$ mesons, mass ratios and numbers of configurations.}
\begin{ruledtabular}
\begin{tabular}{cccccc}
$m_f a$ &
$m_{\pi} a$ &
$m_{\rho} a$ &
$m_{\pi}/m_{\rho}$ &
$m_{a_1}/m_{\rho}$ &
\textrm{$N_{config}$
\footnote{Number of configurations separated by 1000 sweeps.}}\\
\colrule
0.08 & 0.667(1)  & 0.950(2)  & 0.702(2) & 1.480(13) & 3000 \\
0.06 & 0.589(1)  & 0.904(2)  & 0.652(3) & 1.511(19) & 3000 \\
0.04 & 0.503(1)  & 0.861(2)  & 0.584(2) & 1.540(19) & 7964 \\
\end{tabular}
\end{ruledtabular}
\label{kappa}
\end{table}

The propagators of $\pi$, $\rho$, and $a_1$ mesons for $m_f a = 0.08$ are shown in Fig.~\ref{propa}. 
The effective masses, $m_{\rm eff} a$, of these mesons are displayed in Fig.~\ref{eff_mass},
which are determined as 
\begin{eqnarray}
 \frac{G (t)}{G(t+1)} & = & 
 \frac{e^{-m_{\rm eff}(t)t} +  e^{-m_{\rm eff}(t)(T-t) }}{e^{-m_{\rm eff}(t)(t+1)} +  e^{-m_{\rm eff}(t)(T-(t+1)) }} \ , \ 
\end{eqnarray}
where $G(t)$ represents the propagators of the mesons. 
We estimate the statistical errors using the jackknife method.
Thanks to the large enough statistics, we obtain very clear propagators and effective masses for the $a_1$ meson. 
The masses of the $\pi$, $\rho$, and $a_1$ mesons for $m_f a = 0.80$, 0.06, and 0.04 are listed in Table~\ref{kappa}. 
The $\pi$ and $\rho$ masses are evaluated from effective masses in the range of $6 \le t/a \le 9$. 
The $a_1$ mass, on the other hand, is obtained  in the range of $5 \le t/a \le 8$,   
because the effective masses of $a_1$ suffer from large errors at large $t$.

\begin{figure}[h]
\begin{center}
\includegraphics[scale=0.57]{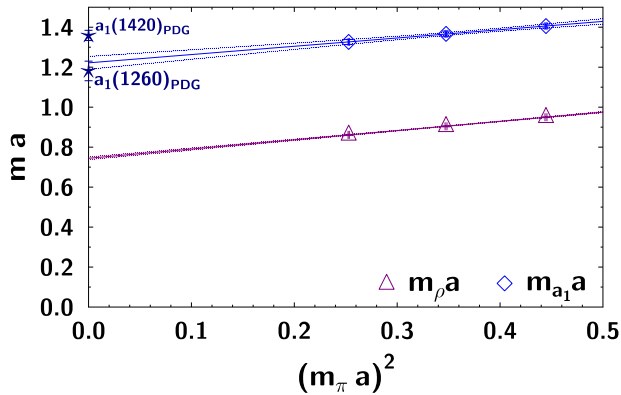}
\caption{(color online). 
Dependences of 
$\rho$ meson masses $m_{\rho}a$ (open triangles) 
and $a_1$ meson masses $m_{a_1}a$ (open diamonds)
on $(m_\pi a)^2$.
Lines for $m_{\rho}a$ and $m_{a_1} a$ show linear fits. 
Stars represent the experimental values of $a_1$(1260) and $a_1$(1420)~\cite{PDG2018}. 
}
\label{mas_dep2}
\end{center}
\end{figure}

In Table~\ref{kappa}, 
the results for meson masses and mass ratios are summarized, 
while 
those of Blum {\it et al.}~\cite{Blum:2000kn} are shown in Table~\ref{kappa2}. 
The masses of $\pi$ and $\rho$ mesons obtained in our simulation on a small lattice 
show good agreement with those on a large lattice ($8^3 \times 32$), 
though our results are less than 2 percent higher than their results. 
Figure~\ref{mas_dep2} shows that $m_\rho a$ and $m_{a_1} a$ vary linearly with $(m_\pi a)^2$.

\begin{table}%[h]
\caption{
Masses of $\pi$ and $\rho$ mesons, mass ratio and number of configurations 
reported by Blum {\it et al.}~\cite{Blum:2000kn}. 
Simulation parameters are $\beta=5.7$, $8^3 \times 32$, $N_5=32$, and $m_5=1.65$. 
}
\begin{ruledtabular}
\begin{tabular}{ccccc}
$m_f a$ &
$m_{\pi} a$ &
$m_{\rho} a$ &
$m_{\pi}/m_{\rho}$ &
\textrm{$N_{config}$
}\\
\colrule
0.06 & 0.595(9)  & 0.92(2)  & 0.65(2) & 94 \\
0.04 & 0.502(5)  & 0.87(4)  & 0.58(3) & 184 \\
\end{tabular}
\end{ruledtabular}
\label{kappa2}
\end{table}

In the chiral limit, $(m_\pi a)^2 = 0$. 
Using $m_{\rho} = 775$\ MeV as the input, we obtain $a=0.190(2)$\ fm. 
In this limit, the difference between the chiral extrapolations $m_{f}a \rightarrow 0$ and $m_{f}a \rightarrow -m_{\rm res} a$ 
is negligible due to the smallness of $m_{\rm res} a  =1.27\times10^{-2}$, where $m_{\rm res}$ is the residual mass. 
Therefore, we apply $m_{f} a \rightarrow 0$. 
We estimate the mass ratio $m_{a_{1}}/m_{\rho}$ to be 1.64(6) and the mass of the $a_1$ meson to be $m_{a_1} = 1272(45)$\ MeV. 
Our result is consistent with the experimental value of 1230(40)\ MeV~\cite{PDG2018}.

%\section{Conclusion and discussion}
\noindent\underline{\bf Conclusion and Discussion} 
\vspace{2mm}

We studied the lightest $a_1$ meson based on a quenched approximation using truncated overlap fermions. 
We estimated the mass of the $a_1$ meson to be 1272(45)\ MeV, which is 
in good agreement with the experimental value for $a_1$(1260)~\cite{PDG2018}. 
The masses obtained in our simulation should be considered as the upper limits.  
Our results are consistent with those of Wingate {\it et al.}~\cite{Wingate:1995hy} 
who employed a full QCD simulation without chiral symmetry. 
Our simulation used truncated overlap fermions, and thus respects chiral symmetry, 
but in the quench approximation. 

Gattringer {\it et al.} determined the mass of the $a_1$ meson 
using the chirally improved Dirac operator in the quenched approximation 
with the L{\"u}scher-Weisz gauge action~\cite{Gattringer:2008}. 
The ground state of $a_1$ meson in their calculation is close to $a_1$(1420). 
Possible reason for the difference between our result and theirs 
is the difference of statistics: our statistics are 30 or 80 times as large as theirs. 
We succeeded in obtaining the lowest state of $a_1$ meson, in spite of utilizing a simple 
two-quark interpolator. 

Our lattice study and quark model analysis support that 
the simple two-body constituent-quark structure of $a_1$(1260) is consistent with the experimentally observed $a_1$(1260). 
Our $a_1$ meson does not agree with $a_1$(1420). 
A quench simulation is a clean theoretical experiment in which virtual intermediate states such as $qq\bar{q}\bar{q}$ are highly suppressed. 
Therefore, $a_1$(1420) may contain an unconventional state, such as $qq\bar{q}\bar{q}$. 
In the $qq\bar{q}\bar{q}$ case, 
dynamical quarks may play an essential role. 
Note that there have been arguments to consider $a_1$(1420) as 
a dynamical effect of the triangle diagram
~\cite{Mikhasenko, Aceti, Basdevant, Wang:2016}. 
Also, $a_1$(1640) might be a radial excitation of $a_1$(1260), according to the quark model analysis. 
We leave it to the future task to complete $a_1$ meson spectroscopy with the lattice QCD simulation.

%\begin{acknowledgments}
\begin{center}
\noindent{\bf acknowledgments} 
\end{center}
This work could not be completed without valuable advices by T.~Kunihiro. 
This work was completed with the support of 
RSF grant 15-12-20008. 

It was also supported in part by 
the JSPS KAKENHI Grant-in-Aid for Scientific Research (S) Grant 
Number JP26220707, the JSPS KAKENHI Grant-in-Aid for Scientific 
Research (C) Grant Number JP17K05438, 
Research Activity of Matsumoto University (No.14111048, No.16111048), 
and 
Scientific Research (Kakenhi) Numbers 24340054 and 26610072. 
The simulation was performed on an NEC SX-ACE supercomputer at RCNP and the Cybermedia Center, Osaka University, 
and was conducted using the Fujitsu PRIMEHPC FX10 System (Oakleaf-FX, Oakbridge-FX) 
at the Information Technology Center, The University of Tokyo. 
This work was supported by 
``Joint Usage/Research Center for Interdisciplinary Large-scale Information Infrastructures" 
in Japan (Project ID: EX17706 and jh180053-NAJ). 
%\end{acknowledgments}


\end{document}